\documentclass[journal]{IEEEtran}

\usepackage{cite}
\usepackage{graphicx}
\graphicspath{{Figures/}} 
\usepackage{amsmath}
\usepackage{amssymb}
\usepackage{algorithmic}
\usepackage{array}
\usepackage{url}
\usepackage{eqparbox}
\usepackage{nomencl}
\usepackage{etoolbox}
\usepackage{booktabs}
\usepackage{enumitem,kantlipsum}
\usepackage{comment}
\usepackage{xcolor}
\usepackage{hyperref}
\hypersetup{
    colorlinks=true,
    linkcolor=blue!80!black,
    citecolor=blue!80!black,
    filecolor=blue!80!black,
    urlcolor=blue!80!black,
    linkbordercolor=blue!80!black,
    citebordercolor=blue!80!black,
    urlbordercolor=blue!80!black
}

\begin{document}
\title{Advanced Resilience Planning for\\ Distribution Systems}
\author{
    \IEEEauthorblockN{
        Ahmad~Bin~Afzal, 
Nabil~Mohammed, Shehab~Ahmed, Charalambos~Konstantinou
    }\\
    \IEEEauthorblockA{CEMSE Division, King Abdullah University of Science and Technology (KAUST)}\\
    \IEEEauthorblockA{E-mail: \{firstname.lastname\}@kaust.edu.sa}
}

\maketitle
\begin{abstract}
Climate change has led to an increase in the frequency and severity of extreme weather events, posing significant challenges for power distribution systems. In response, this work presents a planning approach in order to enhance the resilience of distribution systems against climatic hazards. The framework systematically addresses uncertainties during extreme events, including weather variability and line damage. Key strategies include line hardening, backup diesel generators, and sectionalizers to strengthen resilience. We model spatio-temporal dynamics and costs through a hybrid model integrating stochastic processes with deterministic elements. A two-stage stochastic mixed-integer linear approach is developed to optimize resilience investments against load loss, generator operations, and repairs. Case studies on the IEEE 15-bus benchmark system and a realistic distribution grid model in Riyadh, Saudi Arabia demonstrate enhanced system robustness as well as cost efficiency of 10\% and 15\%, respectively.

\end{abstract}


\makenomenclature
\renewcommand{\nomname}{Nomenclature}

\renewcommand\nomgroup[1]{%
  \item[\itshape
  \ifstrequal{#1}{P}{Variables}{%
  \ifstrequal{#1}{N}{Sets and indices}{%
  \ifstrequal{#1}{O}{Parameters}{}}}%
]}
\nomenclature[N]{$\Omega_{\text{B}}$}{Set of line indices (i, j).}
\nomenclature[N]{$\Omega_{\text{K}}$}{Set of hardening pole type indices k.}
\nomenclature[N]{$\Omega_{\text{L}}$}{Set of loads indices i.}
\nomenclature[N]{$\Omega_{\text{N}}$}{Set of nodes indices i.}
\nomenclature[N]{\(S\)}{Set of sampled scenario indices s.}
\nomenclature[N]{$\mathcal{T}_H$}{Time duration set of climatic hazard indices $t$.}
\nomenclature[O]{$c^{c}_{ij}$}{Cost of installing sectionalizer at the line (i, j).}
\nomenclature[O]{$c^{h}_{ij,k}$}{Cost of hardening line (i, j) with k-th pole type.}
\nomenclature[O]{$c^{g}_i$}{Cost of installing a DG at bus i.}
\nomenclature[O]{$c^{L}_i$}{Cost of shedding 1kWh of i-th load.}
\nomenclature[O]{$c^{o}_i$}{Cost of operating DG at bus i.}
\nomenclature[O]{$P^{L,s}_{i,t}$, $Q^{L,s}_{i,t}$}{Stochastic parameter indicating active/reactive load demand at time t.}
\nomenclature[O]{$P^{\text{max}}_{ij,t}$, $Q^{\text{max}}_{ij,t}$}{Maximum active/reactive line flow at time t.}
\nomenclature[O]{$P^{g,\text{max}}_i$, $Q^{g,\text{max}}_i$}{Active/reactive power limits of DG.}
\nomenclature[O]{$R^{e}_{ij}$, $X^{e}_{ij}$}{Resistance/reactance of line (i, j).}
\nomenclature[O]{$V_0$}{Reference voltage magnitude.}
\nomenclature[O]{$V^{\text{max}}_i$, $V^{\text{min}}_i$}{Maximum/minimum voltage magnitude.}
\nomenclature[O]{$w_H$}{Total occurrence of climatic hazards in a year.}
\nomenclature[O]{$\chi^{s}_{ij,k}$}{Stochastic parameter indicating repair cost of the line (i, j) with k-th pole type.}
\nomenclature[O]{$x^{co}_{ij,i}$}{Binary parameter indicating whether line (i, j) has an existing sectionalizer (1) or not (0) at the end i.}
\nomenclature[O]{$\zeta^{s}_{ij,k,t}$}{Stochastic parameter indicating status of line (i,j) with k-th pole type, damaged(1) or not (0) at time t.}
\nomenclature[P]{$c^{r,s}_{ij}$}{Repair cost of line (i, j).}
\nomenclature[P]{$P^{s}_{ij,t}$, $Q^{s}_{ij,t}$}{Active/reactive power flow of line (i, j) at time t.}
\nomenclature[P]{$P^{g,s}_{i,t}$, $Q^{g,s}_{i,t}$}{Active/reactive power output of DGs at bus i at time t.}
\nomenclature[P]{$u^{s}_{ij,t}$}{Binary variable indicating whether line (i, j) is damaged (1) or not (0) at time t.}
\nomenclature[P]{$V^{s}_{i,t}$}{Voltage magnitude of bus i at time t.}
\nomenclature[P]{$w^{o,s}_{ij,t}$}{Binary variable indicating whether the line is on (1) or off (0) at time t.}
\nomenclature[P]{$x^{h}_{ij,k}$}{Binary variable indicating whether line (i, j) is hardened with k-th pole type (1) or not (0).}
\nomenclature[P]{$x^{g}_{i}$}{Binary variable indicating whether a new DG is placed at bus i (1) or not (0).}
\nomenclature[P]{$y^{c,s}_{ij,t}$}{Binary variable indicating whether the sectionalizer at the line (i, j) is open (1) or not (0) at time t.}
\nomenclature[P]{$y^{r,s}_{i,t}$}{Load shedding percentage of load at bus i at time t.\\}
\printnomenclature

\section{Introduction}

The reliability and efficiency of electric grid distribution networks are crucial for ensuring uninterrupted power supply from substations to end customers and supporting the growing demands of modern society. In such systems, renewable energy sources (RES) and distributed energy resources (DERs) are increasingly adopted for sustainability and efficiency~\cite{1}. DERs, strategically located near loads and consumers, enhance power supply adequacy. However, the distribution system is at increased risk from cyber-attacks and extreme weather events, exacerbated by climate change, potentially causing irreversible damage and major disruptions~\cite{9340265}.  Notable disruptions include the 2017 Hurricane Maria as well as the 2015 Ukraine blackout caused by a cyber-attack, which illustrates the vulnerability of electrical networks~\cite{9351954}.

Considering the vulnerability of existing electricity delivery systems to extreme events, it is crucial to enhance the resilience of the electrical distribution system. The U.S. National Infrastructure Advisory Council (NIAC) defines power system resilience as the ability to anticipate, plan for, absorb, recover from, and adapt to adverse situations~\cite{4}. Efforts to enhance distribution system resilience are classified into operation-based and planning-based methods. Resilience planning supports load recovery and reinforcement, while operational solutions manage real-time catastrophic events~\cite{2}. This paper focuses on a planning-based approach.

Existing literature has extensively examined the topic of distribution network resilience. In~\cite{11}, a resilience-oriented design (ROD) framework for distribution networks is proposed using scenario-based stochastic programming to enhance robustness against extreme weather, with the mixed integer non-linear programming (MINLP) problems converted to mixed integer linear programming (MILP) for improved computational efficiency. In~\cite{12}, a scenario-based stochastic optimization approach is developed to mitigate hurricane effects by creating a robust distribution system planning model. It integrates hardening strategies and distributed generation (DG) placements using conditional value at risk (CVaR) for the assessment. In~\cite{14}, a resilience planning framework is proposed using a stochastic robust optimization model based on historical weather data, employing a column-and-constraint generation (CCG) algorithm and Monte Carlo simulations for strategic underground line placement.

Existing resilience-centric approaches often prioritize immediate mitigation of adverse weather impacts, neglecting resilience achieved through failure-recovery-cost modeling, overlooking stochastic weather occurrences, diverse enhancement techniques, decision-dependent uncertainties, and dynamic recovery mechanisms. To address these gaps, this paper presents a comprehensive framework that combines deterministic and stochastic modeling to enhance resilience in distribution systems. Key contributions include:
\begin{itemize}
    \item 
    Employing a deterministic causal structure with a hybrid independent stochastic process to model spatio-temporal correlations among uncertainties in ROD problems.
    \item
    Utilizing a two-stage stochastic MILP to optimize resilience strategies by implementing several approaches, considering various uncertainties, and incorporating the entire failure-recovery process, thereby modeling both investment and restoration costs.
    \item
    Validating that the proposed design successfully enhances the resilience of distribution systems while demonstrating cost-effectiveness by applying the framework to an IEEE benchmark system as well as to a realistic distribution grid model in Riyadh, Saudi Arabia.
\end{itemize}
The rest of this paper is structured as follows:
Section~\ref{Sec2} outlines the adopted methodology and mathematical formulations of the ROD problem. Section~\ref{Sec4} presents the numerical results of the validation. Finally, Section~\ref{Sec5} concludes the paper.

\section{Resilience Planning Framework}\label{Sec2}
\begin{figure}
    \centering
    \includegraphics[width=0.48\textwidth]{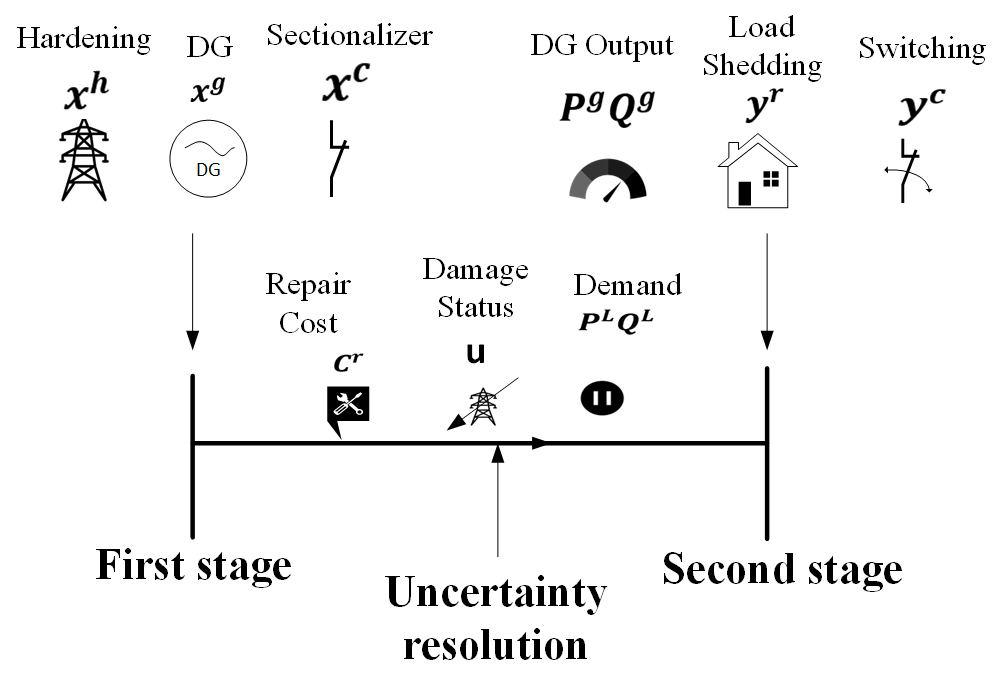}
    \caption{Overall concept of the resilience-driven framework.}
    \label{fig:p2}
\end{figure}

The ROD problem, adopted from\cite{25}, is structured as a two-stage stochastic decision-making process as shown in Fig. \ref{fig:p2}. In the first stage, planners make initial resilience decisions. In the second stage, as the hazard unfolds, operators manage uncertainties, implementing actions like DG re-dispatching, load shedding, and network reconfiguration to minimize costs. Subsection~\ref{subs1} covers decisions and the first-stage problem formulation, Subsection~\ref{subs2} details uncertainties, and Subsection~\ref{subs3} addresses the second-stage problem formulation.

\subsection{First Stage Problem and Decisions}\label{subs1}
In the first stage, decisions are made without knowledge of future scenarios. These include preventive measures such as pole hardening, DG installation, and sectionalizer addition, represented by the decision variable $\mathbf{x} = [\mathbf{x}^h, \mathbf{x}^g, \mathbf{x}^{c_1}]$. Each decision is detailed separately below.
\begin{enumerate}
    \item Hardening Poles: This involves selecting from six different pole class types denoted as:
    \[
    x^{h}_{ij} = [x^h_{ij,1}, x^h_{ij,2}, x^h_{ij,3}, x^h_{ij,4}, x^h_{ij,5}, x^h_{ij,6}],
    \]
    where each component is binary, indicating the selection of a specific pole type for line $(i, j)$. 
    The aggregate decision variables for the network capture pole-hardening decisions across all network lines, 
    \[
    x^h = [x^{h1}, x^{h2}, \ldots, x^{h|\Omega_B|}]
    \]

    \item Installing Backup DGs: The DG placement variable is:
    \[
    x^{g} = [x^{g1}, x^{g2}, \ldots, x^{g|\Omega_N|}], 
    \]
    where each component of \(x^g\) is a binary indicator for the nodes where DGs are positioned.

    \item Sectionalizers:  In the model, installation decisions for sectionalizers are represented by \(x^{c1}_{ij}\), defined as:
    \[
    x^{c1}_{ij} = \begin{bmatrix} x^{c1}_{ij,i} & x^{c1}_{ij,j}\end{bmatrix},
    \]
    where the presence of a sectionalizer at line ends \((i, j)\) is indicated by each element. The array \( \mathbf{x}_{c1} \) details the deployment of sectionalizers across all network line sections in the set \(\{0, 1\}^{2|\Omega_B|}\):
    \[
    x_{c1} = \begin{bmatrix} x^{c1}_1 & x^{c1}_2 & \ldots & x^{c1}_{|\Omega_B|} \end{bmatrix}
    \] 
\end{enumerate}

Based on the preceding decisions, the first-stage problem is formulated as follows:
\begin{equation}
\label{eqf}
\begin{split}
     \min \sum_{(i, j) \in \Omega_{B}} \sum_{k \in \Omega_{K}} c_{i j, k} x_{i j, k}^{h}+
     \sum_{i \in \Omega_{N}} c_{i}^{g} x_{i}^{g} + \\
     \sum_{(i, j) \in \Omega_{B}} c_{i j}^{c}\left(x_{i j, i}^{c_{1}}+x_{i j, j}^{c_{1}}\right)+ 
     \mathcal{Q}(x)
\end{split}
\end{equation}
\begin{equation}
\label{cons1}
s.t. \sum_{k \in \Omega_{K}} x_{i j, k}^{h}=1, \forall(i, j) \in \Omega_{B} 
\end{equation}
\begin{equation}
\label{cons2}
\sum_{i \in \Omega_{N}} x_{i}^{g} \leq N_{G}
\end{equation}
\begin{equation}
\begin{split}
x_{i j, n}^{c_{0}}+x_{i j, n}^{c_{1}}=x_{i j, n}^{c}, \forall(i, j) \in \Omega_{B}, n \in\{i, j\}  \\
x_{i j, k}^{h}, x_{i j, n}^{c_{1}}, x_{i j, n}^{c}, x_{i}^{g} \in\{0,1\}, \forall i \in \Omega_{N},(i, j) \in \Omega_{B}
\end{split}
\end{equation}

\begin{equation}
\label{eq6}
\mathcal{Q}(x)=w_{H} \cdot {E}_{\boldsymbol{\xi}} \phi(x,\boldsymbol{\xi}) \cong w_{H} \cdot \sum_{s \in \mathcal{S}} p_{r}(s) \phi(x, s)
\end{equation}
where \(w_H\) represents the annual count of wind-induced hazards, and \(pr(s)\) is the probability of each scenario, defined as \(\frac{1}{|S|}\). 
The objective function in \eqref{eqf} minimizes costs from first-stage decisions and expected second-stage costs. Eq.~\eqref{cons1} ensures only one hardening strategy per line, while \eqref{cons2} limits DG installations. The presence of a sectionalizer at terminal \(n\) of line \((i, j)\) is indicated by \(x^{c}_{ij,n}\), impacting the second stage. The term \(\mathcal{Q}(\mathbf{x})\) in \eqref{eqf} represents the expected costs from the second stage, incorporating \(\mathbf{x}\) and associated wind-induced hazards.

\subsection{Uncertainty Analysis}\label{subs2}
A hybrid method combining a deterministic causal framework with independent stochastic processes is presented to address uncertainties. The uncertain parameters are explained briefly below:
\begin{enumerate}
    \item \textit{Line Damage Status, \(u(t; x^h)\)}: This uncertainty concerns the operational status of a power line during climatic hazards, whether it remains functional or becomes damaged. The binary variable \(x^h\) directly influence the likelihood of line damage under stressful conditions.
    \item \textit{Repair Cost, \(c^r\)}: 
    This parameter reflects financial uncertainties in repairing damaged infrastructure within the power distribution network. For each line segment \( (i, j) \), six potential cost scenarios are outlined corresponding to different pole types, each influenced by the type of pole used and the severity of resultant damage.
    \item \textit{Power Demand of Load, ($P^{L}, Q^{L}$)}:
    Fluctuations in end users' electrical power consumption during climatic events introduce uncertainties in both active and reactive power demand and impose strain on the system. These uncertainties are modeled by perturbing original system data values, extracting relevant information, and introducing variability within a predefined perturbation range (e.g., 30\%). 
\end{enumerate}

\subsection{Second Stage Problem}\label{subs3}
After an extreme weather event occurs, the system responds to this “realized uncertainty'' with recourse actions including DG re-dispatch, load shedding, and network reconfiguration. Second-stage decisions are represented by vector $y^{R,s} = [P^{g,s}, Q^{g,s}, y^{c,s}, y^{r,s}]$. Sectionalizers are crucial in isolating compromised network segments to minimize service interruptions, aligning with first-stage decision variables $x_{ij,n}^{c}$.

\begin{table}[t]
\centering
\caption{Evaluation of the On-Off Status Variable.}
\label{tab1}
\begin{tabular}{ccc|ccc}
\midrule
\toprule
\(x^c_{ij,n}\) & \(y^{c,s}_{ij,t}\) & \(w^{o,s}_{ij,t}\) & \(x^c_{ij,n}\) & \(y^{c,s}_{ij,t}\) & \(w^{o,s}_{ij,t}\) \\
\hline
0 & 0 & 1 & 1 & 0 & 1 \\
0 & 1 & N/A & 1 & 1 & 0 \\
\hline
\bottomrule
\end{tabular}
\\
\textit{*N/A: the case should be infeasible.}
\end{table}

\subsubsection{Cost Objective Function}
The primary goal is to ensure that the system's response to the realized scenario is cost-effective and efficient in mitigating its impact.
\begin{equation}
\begin{aligned}
\label{eq7}
\phi(\boldsymbol{x},s)=\min\sum_{i\in\Omega_{N}} \sum_{t\in\mathcal{T}_{H}^{s}}c_{i}^{L} y_{i, t}^{r, s} P_{i, t}^{L, s}\Delta t \\ +\sum_{i\in\Omega_{N}}\sum_{t\in\mathcal{T}_{H}^{s}} c_{i}^{o} P_{i, t}^{g, s}\Delta t & + \sum_{(i, j) \in \Omega_{B}} c_{ij}^{r, s} 
\end{aligned}
\end{equation}

\subsubsection{Line Damage Status Constraint}
\begin{equation}
u_{i j, t}^{s}=\sum_{k \in \Omega_{K}} x_{i j, k}^{h} \zeta_{i j, k, t}^{s}, \forall(i, j) \in \Omega_{B}, t \in \mathcal{T}_{H}^{s}
\end{equation}
\subsubsection{Line Repair Cost Constraint}
\begin{equation}
c_{i j}^{r, s}=\sum_{k \in \Omega_{K}} x_{i j, k}^{h} \chi_{i j, k,}^{s}, \forall(i, j) \in \Omega_{B}
\end{equation}

\subsubsection{Line’s On-Off Status Constraints}
The variable $x_{ij,n}^{c}$ denotes the line's on-off state for nodes $\forall n \in \{i, j\}, (i, j) \in \Omega_{B}$. Introducing $w_{ij,t}^{o,s}$ for $\forall (i, j) \in \Omega_{B}, t \in T_s^H$ effectively captures this state, with Table~\ref{tab1} specifying optimal values based on combinations of $x_{ij}^{c}$ and $y_{ij,t}^{c,s}$.
\begin{equation}
y_{i j, t}^{c, s} \leq x_{i j}^{c}, \forall(i, j) \in \Omega_{B}, t \in \mathcal{T}_{H}^{s}
\end{equation}
\begin{equation}
x_{i j}^{c}+y_{i j, t}^{c, s}+2 w_{i j, t}^{o, s} \geq 2, \forall(i, j) \in \Omega_{B}, t \in \mathcal{T}_{H}^{s}
\end{equation}
\begin{equation}
w_{i j, t}^{o, s}+y_{i j, t}^{c, s} \leq 1, \forall(i, j) \in \Omega_{B}, t \in \mathcal{T}_{H}^{s} 
\end{equation}
\begin{equation}
y_{i j, t}^{c, s}, w_{i j, t}^{o, s} \in\{0,1\}, \forall(i, j) \in \Omega_{B}, t \in \mathcal{T}_{H}^{s}
\end{equation}
\subsubsection{Line Flow Limits}
\begin{equation}
-w_{i j, t}^{o, s} P_{i j}^{\max } \leq P_{i j, t}^{s} \leq w_{i j, t}^{o, s} P_{i j}^{\max }, \forall(i, j) \in \Omega_{B}, t \in \mathcal{T}_{H}^{s} 
\end{equation}

\begin{equation}
-w_{i j, t}^{o, s} Q_{i j}^{\max } \leq Q_{i j, t}^{s} \leq w_{i j, t}^{o, s} Q_{i j}^{\max }, \forall(i, j) \in \Omega_{B}, t \in \mathcal{T}_{H}^{s} 
\end{equation}

\subsubsection{Linearized DistFlow Equations}
\begin{equation}
\begin{aligned}
\sum_{\{j \mid(i, j) \in \Omega_{B}\}}P_{i j,t}^{s}=P_{i, t}^{g, s}-\left(1-y_{i, t}^{r, s}\right) P_{i, t}^{L}-\\\varepsilon_{1} V_{i, t}^{s}, \forall i \in \Omega_{N}, t \in \mathcal{T}_{H}^{s}
\end{aligned}
\end{equation}

\begin{equation}
\begin{aligned}
\sum_{\left\{j \mid(i, j) \in \Omega_{B}\right\}} Q_{i j, t}^{s}=Q_{i, t}^{g, s}-\left(1-y_{i, t}^{r, s}\right) Q_{i, t}^{L},\\ \forall i \in \Omega_{N}, t \in \mathcal{T}_{H}^{s}
\end{aligned}
\end{equation}

\begin{equation}
\begin{aligned}
V_{i, t}^{s} &-\frac{R_{i j}^{e} P_{i j, t}^{s}+X_{i j}^{e} Q_{i j, t}^{s}}{V_{0}}-\left(1-w_{i j, t}^{o, s}\right) M_{1} \leq V_{j, t}^{s} \leq V_{i, t}^{s}\\ &-\frac{R_{i j}^{e} P_{i j, t}^{s}+X_{i j}^{e} Q_{i j, t}^{s}}{V_{0}}+\left(1-w_{i j, t}^{o, s}\right) M_{1},\\ & \forall i \in \Omega_{N}, t \in \mathcal{T}_{H}^{s}
\end{aligned}
\end{equation}

\subsubsection{Voltage Magnitude Limits}
An auxiliary binary variable \( w_{i,t}^{m,s} \) introduces alternative voltage magnitude limits: setting \( V_{i,t}^{s} \) to zero when \( w_{i,t}^{m,s} = 0 \) and within a safe range when \( w_{i,t}^{m,s} = 1 \). Constraints \eqref{eq21}-\eqref{eq22} ensure zero voltage magnitude at the node if line $(i, j)$ is destroyed.
\begin{equation}
\label{eq21}
w_{i, t}^{m, s} V_{i}^{\min } \leq V_{i, t}^{s} \leq w_{i, t}^{m, s} V_{i}^{\max }, \forall i \in \Omega_{N}, t \in \mathcal{T}_{H}^{s}
\end{equation}
\begin{equation}
\label{eq22}
u_{i j, t}^{s}+w_{i, t}^{m, s} \leq 1, \forall(i, j) \in \Omega_{B}, \forall i \in \Omega_{N}, t \in \mathcal{T}_{H}^{s}
\end{equation}
\begin{equation}
w_{i, t}^{m, s} \in\{0,1\}, \forall i \in \Omega_{N}, t \in \mathcal{T}_{H}^{s}
\end{equation}

\subsubsection{Load Shedding Ratio Limit}
Constraint (\ref{eq:load-shedding-limit}) establishes the load-shedding ratio limit. At a node where the voltage magnitude is zero, the load shedding ratio will be 1. The cost of load shedding serves as a severity index for climate threats in the objective function.
\begin{equation}\label{eq:load-shedding-limit}
1-w_{i, t}^{m, s} \leq y_{i, t}^{r, s} \leq 1, \forall i \in \Omega_{N}, t \in \mathcal{T}_{H}^{s}
\end{equation}       

\subsubsection{DG Capacity Limits and Status}
When a DG is installed at any node, the following constraints are applied:
\begin{equation}
0 \leq P_{i, t}^{g, s} \leq x_{i}^{g} P_{i}^{g, \max }, \forall i \in \Omega_{N}, t \in \mathcal{T}_{H}^{s} 
\end{equation}      
\begin{equation}
0 \leq Q_{i, t}^{g, s} \leq x_{i}^{g} Q_{i}^{g, \max }, \forall i \in \Omega_{N}, t \in \mathcal{T}_{H}^{s}
\end{equation}

\section{Results}\label{Sec4}
\begin{figure}
    \centering
    \includegraphics[width=0.4\textwidth]{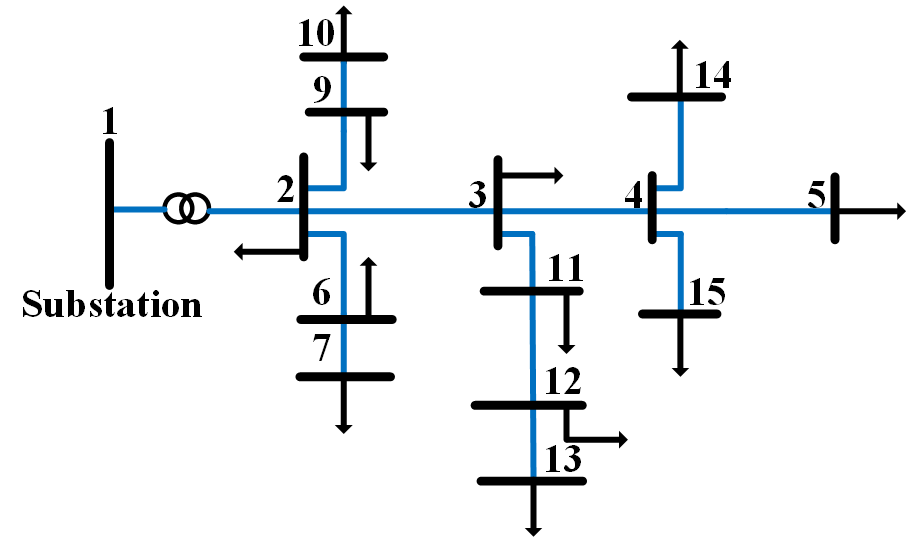}
    \caption{IEEE 15-bus distribution system.}
    \label{fig:p5}
\end{figure}

\begin{figure}[t]
    \centering
    \includegraphics[width=0.4\textwidth]{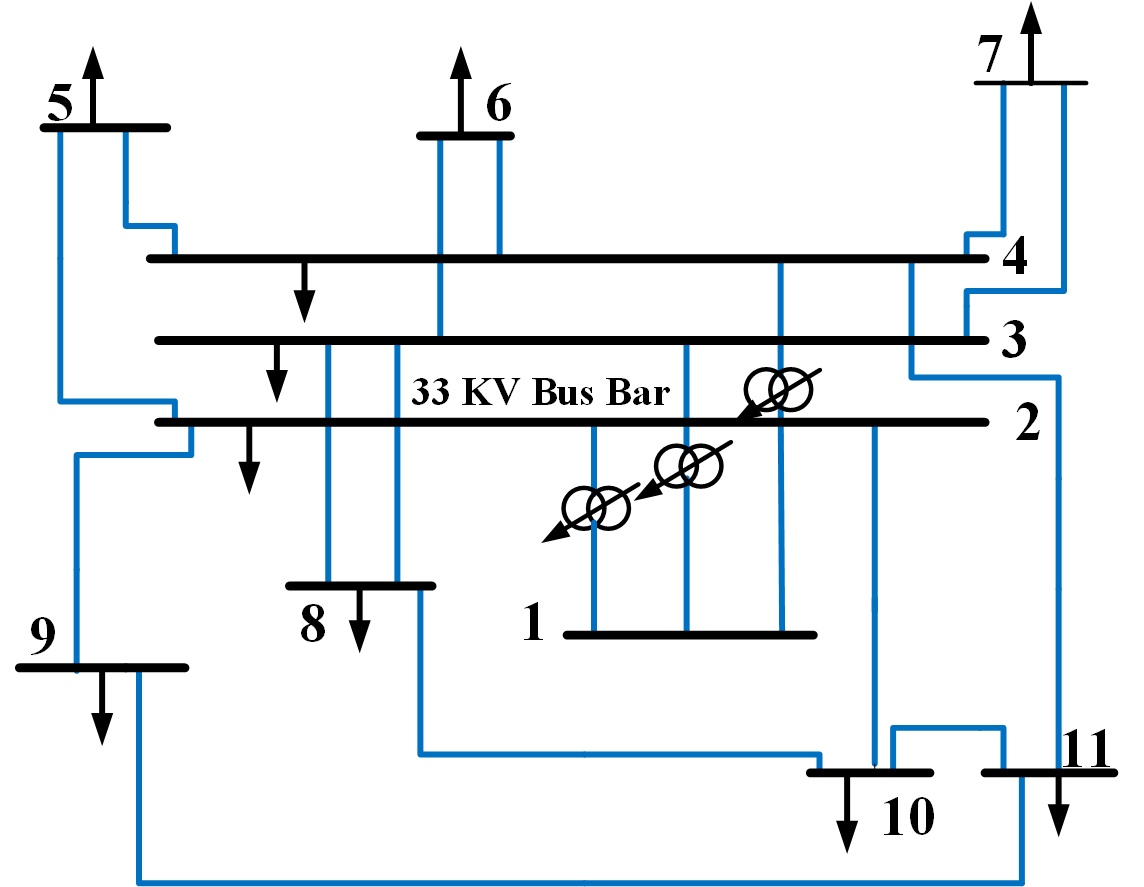}
    \caption{Saudi Consolidated Electricity Central Region Riyadh, Saudi Arabia, adopted from \cite{26}.}
    \label{fig:p14}
\end{figure}

Numerical studies are conducted using the IEEE 15-bus distribution system and a realistic distribution grid in Riyadh, Saudi Arabia as shown in Fig.~\ref{fig:p5} and Fig.~\ref{fig:p14}, respectively. The use of the Saudi Electrical Company (SEC) system in this research highlights the practical applicability and effectiveness of the framework. By applying the suggested techniques to the SEC network, the study demonstrates the translation of theoretical models into real-world scenarios. Furthermore, utilizing both the IEEE 15-bus and SEC systems as case studies showcases the framework's flexibility across various configurations. 

The capital costs for the ROD method are outlined in Table~\ref{tab:table_2}. A basic load-shedding cost of \$14/kWh is assumed. The repair cost for the six pole types is assumed uniform (\(\chi^{p}_{ij,1} = \dots = \chi^{p}_{ij,6} = \$4000\)). The operation cost of backup DGs is set at \$8/kWh. A time step of \(\Delta t = 2\) hours was used. To evaluate system resilience, 10 scenarios are randomly generated, each representing a range of potential disturbances and with a probability \(p_{r}(s) = 1/10\). The dual-stage optimization problem was formulated and solved using the Pyomo module in Python, with the MIP solver CPLEX and in conjunction with the mpi-sppy module efficiently handling different scenarios.

\begin{table}[t]
\centering
\caption{Investment cost of different ROD methods.}
\label{tab:table_2}
\begin{tabular}{ccc}
\midrule
\toprule
\#No. & Methods & Cost (\$) \\[1.6ex] \hline
1 & Upgrading pole class type 1 & 10000/pole\\
2 & Upgrading pole class type 2 & 20000/pole\\
3 & Upgrading pole class type 3 & 30000/pole\\
4 & Upgrading pole class type 4 & 15000/pole\\
5 & Upgrading pole class type 5 & 25000/pole\\
6 & Upgrading pole class type 6 & 35000/pole\\
7 & Installing a DG with 400kW capacity & 1,000/kW\\
8 & Adding an automatic sectionlizer & 15,000 \\ \hline
\bottomrule
\end{tabular}
\end{table}

\subsection{Combined Objective Cost Analysis}
Table~\ref{tab3} illustrates the total costs incurred in both systems across ten scenarios, combining preventive measures in the first stage and corrective actions in the second. The values of the objective function show unpredictability due to intrinsic uncertainties from random variables in the simulation.
The results reveal significant discrepancies in total objective costs between the IEEE 15-bus system and the SEC system. For instance, scenario S5 highlights this difference, with costs of \$805,119.63 and \$471,717.23, respectively. This disparity suggests potential intrinsic benefits of the SEC system's topology in terms of robustness and cost-effectiveness under certain conditions. Lower costs, influenced by uncertainties like load demand tied to system data, indicate more stable or ideal conditions, reflecting calculated characteristics and reduced deviation from system ratings.

\begin{table}[b]
\centering
\caption{Combined objective cost values for each scenario in the SEC system and IEEE 15 bus system (in \$).}
\label{tab3}
\begin{tabular}{ccc}
\midrule
\toprule
Scenario & SEC (\$) & 15-bus (\$) \\[1.6ex]
\hline
S1 & 562219.63 & 733092.48\\
S2 & 570513.04 & 699109.10\\
S3 & 489183.24 & 672546.83\\
S4 & 488079.29 & 601663.27\\
S5 & 471717.23 & 805119.63\\
S6 &  473673.49 & 697736.15\\
S7 & 491337.31 & 717141.86\\
S8 & 510199.15 & 610958.30\\
S9 & 473137.23 & 743544.95\\
S10 & 494909.45 & 684406.91 \\ 
\hline
\bottomrule
\end{tabular}
\end{table}

\subsection{Load Shedding Cost Comparison with and without ROD}
\begin{figure}[t]
    \centering\includegraphics[width=0.45\textwidth]{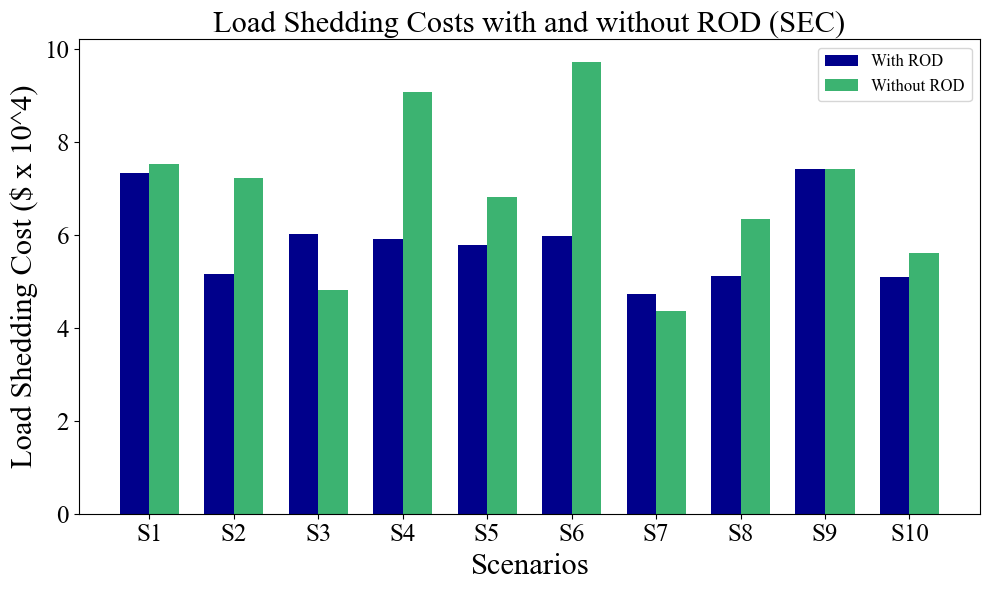}
    \caption{Load shedding cost comparison with and without ROD under different scenarios on the SEC system. }
    \label{fig:p10}
\end{figure}

\begin{figure}[t]
    \centering
    \includegraphics[width=0.45\textwidth]{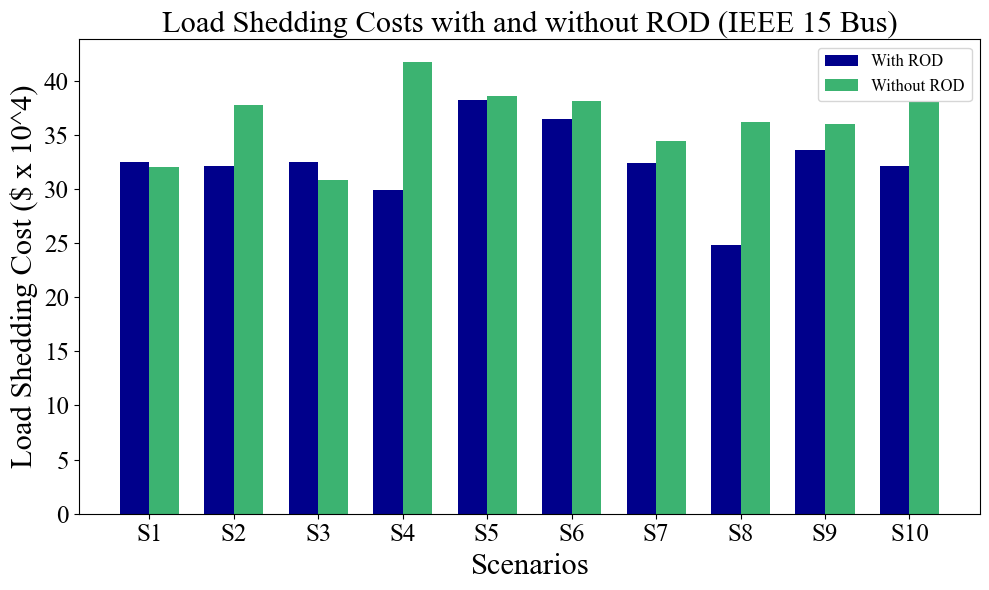}
    \caption{Load Shedding cost comparison with and without ROD under different scenarios on the IEEE 15 bus system. }
    \label{fig:p11}
\end{figure}

Fig.~\ref{fig:p10} and Fig.~\ref{fig:p11} depict the comparison of load-shedding costs with and without ROD in the IEEE 15-bus and SEC distribution systems, respectively. These costs are crucial in this resilience study, impacting economic losses, power reliability, and vital services continuity. Scenarios are generated to simulate different disturbance levels affecting grid operations, leading to vulnerable line outages. Decreased load-shedding expenses are consistently demonstrated by the implementation of ROD. In scenario 6 of Fig. 4, for example, the ROD approach shows a cost difference of nearly 40\% compared to the non-ROD method. Average cost savings in the SEC system approach 15\%, while in the IEEE 15-bus system (Fig. 5), scenario S4 indicates a close to 10\% reduction in costs with ROD, highlighting its resilience benefits.  The effectiveness of ROD is attributed to early-stage planning actions such as sectionalizers, DG placement, and line hardening.

\section{Conclusion}\label{Sec5}

This paper outlines a resilience approach for safeguarding power distribution systems from wind-induced climate hazards. Strong spatial and temporal correlations are demonstrated between ROD decisions, uncertainty space, and system operations during and after climatic disasters. To minimize investment costs initially and predict load loss, repair, and DG operation costs subsequently, a two-stage stochastic mixed-integer model is presented. Numerical studies on the IEEE 15-bus distribution system and the SEC system validate the effectiveness of optimal ROD in enhancing system resilience. 15\% in the SEC system and 10\% in the IEEE 15-bus system, highlighting its significant impact on reducing load-shedding costs and enhancing overall system resilience. Future research could explore larger networks and examine the post-hazard status of the system. 

\section*{Acknowledgment}
This publication is based upon work supported by King Abdullah University of Science and Technology (KAUST) under Award No. ORFS-CRG11-2022-5021.

\bibliographystyle{ieeetr}
\bibliography{reference}
\vfill
\end{document}